\documentclass[aps,amssymb,amsmath,singlecolumn,prb,superscriptaddress]{revtex4}
\usepackage{graphicx}
\usepackage{epsfig}
\usepackage{bm}
\usepackage{bbm}
\usepackage[colorlinks, citecolor=red]{hyperref}
\usepackage{color}
\usepackage[up]{subfigure}

\newcommand{\be}{\begin{equation}}
\newcommand{\ee}{\end{equation}}
\newcommand{\bc}{\begin{center}}
\newcommand{\ec}{\end{center}}
\newcommand{\bea}{\begin{eqnarray}}
\newcommand{\eea}{\end{eqnarray}}
\newcommand{\ba}{\begin{array}}
\newcommand{\ea}{\end{array}}

\begin{document}
\title{Two-component Dirac-like Hamiltonian for generating quantum walk on one-, \\two- and three-dimensional lattices}

\author{C. M. Chandrashekar}
\email{c.madaiah@oist.jp}
\affiliation{Quantum Systems Unit, Okinawa Institute of Science and Technology Graduate University, Okinawa, Japan}

\begin{abstract}
\noindent
From the unitary operator used for implementing two-state discrete-time quantum walk on one-, two- and three- dimensional lattice we obtain a two-component Dirac-like Hamiltonian. In particular, using different pairs of Pauli basis as position translation states we obtain three different form of Hamiltonians for evolution on one-dimensional lattice. We extend this to two- and three-dimensional lattices using different Pauli basis states as position translation states for each dimension and show that the external coin operation, which is necessary for one-dimensional walk is not a necessary requirement for a walk on higher dimensions but can serve as an additional resource to control the dynamics. The two-component Hamiltonian we present here for quantum walk on different lattices can serve as a general framework to simulate, control, and study the dynamics of quantum systems governed by Dirac-like Hamiltonian. 
\end{abstract}

\maketitle
Quantum walks, developed as the quantum analog of the classical random walks\,\cite{Ria58, Fey86, Par88,  LP88, ADZ93, Kem03} first emerged as a powerful tool in the development of quantum algorithms\,\cite{Amb03, CCD03, SKW03, AJR05}. Subsequently, its rich dynamics constituting evolution in superposition of position space has been used as a framework to understand and simulate the dynamics in various systems. For example, they have been used to explain phenomena such as the breakdown of an electric-field driven  system\,\cite{OKA05} and mechanism of wavelike energy  transfer  within photosynthetic systems\,\cite{ECR07, MRL08}, to demonstrate the coherent quantum control over atoms\,\cite{CL08}, localization of Bose-Einstein condensates in optical lattice\,\cite{Cha11a} and to explore topological phases\,\cite{KRB10}. 
The discrete-time version of the quantum walk (DQW)\,\cite{ADZ93, DM96, ABN01, NV01, BCG04, Ken07, CSL08, Rom09, SK10} on a two-state particle is described using a quantum coin operation acting on the internal state of the particle followed by a shift operator to evolve the particle coherently in superposition of different location in position space. During last few years, experimental implementation of the DQW have been demonstrated with energy levels in NMR\,\cite{RLB05}, ions\,\cite{SMS09, ZKG10}, photons\,\cite{PLP08, PLM10, SCP10, BFL10}, and atoms\,\cite{KFC09}. These experimental implementations using one-dimensional (1D) DQW model on a two-level system using a degree two coin operation have opened up a new dimension to simulate quantum dynamics in physical systems like the recent demonstration of localization of photon's wavepacket\,\cite{SCP10}.  Now the immediate interest would be to extend the implementation to two-dimensional (2D) and three-dimensional (3D) lattice structure with the available resources. One of the extension to the 2D is the Grover walk which is defined on a four-level particle with a specific initial state\,\cite{MBS02, TFM03, WKK08}. An alternative extension to  higher ($d$) dimensions is to use a $d$ coupled qubits to describe the internal states\,\cite{ EMB05, OPD06}. This is extremely challenging with the available resources to implement it experimentally. To overcome this challenge, an alternative 2D DQW using a two-state particle resulting in a probability distribution equivalent to the probability distribution from the Grover walk was very recently proposed\,\cite{FGB11, CB13}.  The evolution in one of the scheme\,\cite{FGB11} is composed of evolving the particle in superposition of position space using Hadamard coin operation followed by a shift operation in one of the dimension followed by the evolution in the other dimension with the Hadamard coin operation followed by a shift in that dimension. An evolution in an other two-state scheme is composed of evolving the particle in superposition of position space using the basis states of one of the Pauli operator as the position translation state for one dimension followed by evolving in an other dimension using the basis states of a different Pauli operator as position translation state\,\cite{CB13}.  The DQW on square lattice using both these two-state schemes were shown to be more tolerant to noise when compared to the four-state Grover walk\,\cite{CB13}. In addition, we know that the dynamics in many natural quantum systems are governed by the Hamiltonian. Therefore, it is extremely important to explore the relevant Hamiltonian form to describe the two-state DQW on 1D, 2D, and 3D lattices. This will give way to simulate and explore the possibility of mimicking the dynamics in various naturally occurring physical systems. 
\par
In this report, starting from the unitary operator used to describe the evolution of DQW in 1D with different Pauli basis states as the position translation states, we obtain the Hamiltonian for the evolution on the discrete space and show that its differential operator forms are analogous to the two-component Dirac Hamiltonian. By setting different Pauli basis states as the position translation states for each dimension we show that the two-state DQW  can be realized  on a physically relevant  2D: square, triangular, kagome, and 3D: cubic lattice structures. The Hamiltonian form we obtain for the DQW evolution on square and cubic lattice in the differential operator form are structurally identical to the two-component Dirac Hamiltonian for relativistic particle in two- and three- spatial dimensions. In absence of the quantum coin operation, the Hamiltonian we obtain is identical to the massless case of the relativistic evolution and in presence of the quantum coin operation, the Hamiltonian form for the relativistic evolution of the massive particle is reproduced. Some of the earlier works have established the connection between the DQW and the Dirac equations, by discretizing the Dirac equation to realize unitary cellular automata\,\cite{BB94}, by using the propagator approach in continuum limit for quantum lattice gas automata\,\cite{DM96}, and by analyzing the 1D DQW dynamics in different continuous and small incremental settings at vanishing (very small) value of $\theta$ in the quantum coin operation used for the evolution\,\cite{Str07, CBS10, GDB12}.  In this report, we use the small spatial incremental setting ($z \rightarrow z\pm1$) to arrive at the differential operator form and establish the connection between the DQW and Dirac Hamiltonian for evolution in all three spatial dimensions and all values of $\theta$.  We also show that Hamiltonian form from the evolution operator for DQW on the triangular and kagome lattice is notably different from the Dirac-like Hamiltonian with a second order derivative terms. Operators acting on the basis states of the Pauli matrices are very common in quantum optics experiments and various other physical systems making this scheme of using Pauli basis as position translation states an implementable one in the present experimental setups. We also show that the external coin operation ($\theta \neq \frac{n\pi}{2}$, $n \in {\mathbbm  I}$) which is required for 1D DQW to evolve with an intriguing interference is not a necessary requirement for 2D and 3D DQW using Pauli basis as position translation states. However, an external coin operation can be used as an additional degree of freedom to control the dynamics. Therefore, the DQW using Pauli basis states as position translation states and the two-component Hamiltonian for the evolution on the discrete-position space which is presented in this report, can serve as a general framework to simulate, control, and study the dynamics in different 2D and 3D physical systems.

\section{Hamiltonian for generating one-dimensional DQW}
\label{DQW}
The standard form of DQW evolution on a two-state particle in 1D lattice is defined on a coin (c) and the position (p) Hilbert space ${\cal H} =  {\cal H}_c\otimes    {\cal     H}_p$. The basis states of ${\cal H}_c$ is the position translation states which can be the internal states of the particle, $|\downarrow \rangle = \begin{bmatrix} 1  \\ 0 \end{bmatrix}$ and $|\uparrow \rangle =\begin{bmatrix} 0  \\ 1 \end{bmatrix}$ or any of the pair of the basis states of the Pauli operators. The basis states of ${\cal H}_p$ is described in terms of $| j\rangle$, where $j \in {\mathbbm  I}$, the set of integers associated with the lattice sites. Below, starting from the unitary operator for DQW evolution using different Pauli basis states as the position translation state we obtain the Hamiltonian for walk on discrete-space for each case and show that differential operator form of these Hamiltonians in a finite spatial incremental setting takes the form of the two-component Dirac Hamiltonian.

\vskip 0.2in
\noindent
{\bf Basis states of $\hat{\sigma}_z$ as position translation state~:~} 
For Pauli operator $\hat{\sigma}_z= \begin{bmatrix} 1 & ~~0\\ 0 & -1 \end{bmatrix}$, the eigenstates are $|+\rangle_{\sigma_z} = |\downarrow \rangle$ and $|-\rangle_{\sigma_z} = |\uparrow \rangle$. For the evolution using basis states of Pauli operator $\hat{\sigma}_z$ as position translation state we will choose $|z\rangle$ where $z \in  {\mathbbm  I}$ as the position space. Each step of 1D DQW evolution, $\hat{W}_{\sigma_z} (\theta)$ is described using a  quantum coin operation $\hat{B}_{\sigma_z} (\theta)  \equiv \begin{bmatrix}\cos(\theta)   &   \sin(\theta)  \\ 
-\sin(\theta) &  \cos(\theta) 
\end{bmatrix}$
which evolves the particle (coin) into the superposition of its basis states followed by the unitary shift operator 
\be
\hat{S}_{\sigma_z}\equiv         \sum_z \left [  |\downarrow \rangle\langle
\downarrow|\otimes|z-1\rangle\langle   z|   +  | \uparrow \rangle\langle
\uparrow|\otimes |z+1\rangle\langle z| \right ],
\ee
which shift the state of the particle in superposition of the position space. That is, 
\be
\label{Wop}
\hat{W}_{\sigma_z} (\theta)\equiv \hat{S}_{\sigma_z}[\hat{B}_{\sigma_z}(\theta) \otimes  {\mathbbm 1}] = \sum_{z} \Bigg ( \begin{bmatrix} 
  \cos(\theta)         &     \sin(\theta)
  \\ 0  &   0 
\end{bmatrix}   \otimes |z-1 \rangle \langle z |  
 + \begin{bmatrix}
  0      &          0
  \\ - \sin(\theta) & \cos(\theta) 
\end{bmatrix} \otimes |z+1 \rangle \langle z | \Bigg ).
\ee
The state after the $t$ steps evolution of the DQW is given by,
$|\Psi_t\rangle=[\hat{W}_{\sigma_z}(\theta)]^t|\Psi_{\rm in}\rangle$,
where $|\Psi_{\rm in}\rangle= \left ( \cos(\delta/2)| \downarrow \rangle + e^{i\eta}
\sin(\delta/2)|\uparrow \rangle \right )\otimes |0\rangle$,
is the initial state of the particle at origin, $z=0$. The coin parameter $\theta$ in $\hat{W}_{\sigma_z}(\theta)$ controls the variance of the probability distribution of the walk. Equivalently,  the initial state of the particle at the origin can be written as a two-component wavefunction, $|\Psi_{in}\rangle = \begin{bmatrix} 
  \psi^{\downarrow}(0, 0) \\
  \psi^{\uparrow}(0, 0) \\
  \end{bmatrix}$  and the net evolution operator for implementing each step of the DQW will be,
  \bea
 \label{paulizshif}
\hat{W}_{\sigma_z}(\theta)
=  \begin{bmatrix}
   \mbox{~~}\cos(\theta)e^{-i\hat{P}_z}      &     &     \sin(\theta)e^{-i\hat{P}_z}
  \\ - \sin(\theta)e^{+i\hat{P}_z } & & \cos(\theta)e^{+i\hat{P}_z  } 
\end{bmatrix}. 
\eea 
Here $\hat{P}_z$ is the position displacement operator whose action on all position $z$ is local such that $e^{\pm i \hat{P}_z }\psi^{\downarrow (\uparrow)}(z, t) = \psi^{\downarrow (\uparrow)}(z\pm 1, t)$\,\cite{ADZ93, Kem03}. The state of the particle after any time $t$ ($t$ steps) will be
\bea
|\Psi (t ) \rangle = [\hat{W}_{\sigma_z}(\theta)]^t  \begin{bmatrix} 
  \psi^{\downarrow}(0, 0) \\
  \psi^{\uparrow}(0, 0) \\
  \end{bmatrix} =    \sum_z \begin{bmatrix} 
  \psi^{\downarrow}(z, t) \\
  \psi^{\uparrow}(z, t) \\
  \end{bmatrix},
\eea
and the unitary operator  $\hat{W}_{\sigma_z}(\theta) \equiv e^{-i\hat{H}_{\sigma_z}(\theta)}$. We have taken $\hbar = 1$ and $\hat{H}_{\sigma_z}(\theta)$ is the effective Hamiltonian corresponding to the unitary operator $\hat{W}_{\sigma_z}(\theta)$.  By taking the logarithm of $\hat{W}_{\sigma_z}(\theta)$, the effective form of the time independent  Hamiltonian we obtain for the quantum walk on discrete position space is,
\bea
\label{hamil2a}
\hat{H}_{\sigma_{z}}(\theta)=  \frac{\hat{\omega}_z}{\sin(\hat{\omega}_z)} 
\begin{bmatrix}
\cos(\theta) \sin(\hat{P}_z)  & 
-i\sin(\theta)e^{-i\hat{P}_z} \\
-i \sin(\theta)e^{i\hat{P}_z} 
&
\cos(\theta) \sin(\hat{P}_z) 
\end{bmatrix} \cdot \hat{\sigma}_z.
\eea
Here $\hat{\omega}_z = \cos^{-1}\Big (\cos(\theta)\cos(\hat{P}_z)\Big)$ and the action of $\sin(\hat{P}_z)$ and $\cos(\hat{P}_z)$ on state $\psi^{\downarrow (\uparrow)}(z, t)$
is given by 
\bea
\sin(\hat{P}_z) \psi^{\downarrow (\uparrow)}(z, t) &= \frac{i}{2} \big [ \psi^{\downarrow (\uparrow)}(z-1, t) - \psi^{\downarrow (\uparrow)}(z+1, t) \big ], \\
\cos(\hat{P}_z) \psi^{\downarrow (\uparrow)}(z, t) &=   \frac{1}{2} \big[ \psi^{\downarrow (\uparrow)}(z-1, t) + \psi^{\downarrow (\uparrow)}(z+1, t) \big ]. 
\eea
The differential operator form of the Eq.\,(\ref{hamil2a}) will be,
\bea
\label{hamil2aa}
\hat{H}_{\sigma_{z}}(\theta)=  -i  \begin{bmatrix}
\cos(\theta)   & 
~~\sin(\theta) \\
 \sin(\theta) 
&
- \cos(\theta) 
\end{bmatrix}   \cdot \frac{\partial}{\partial z} + \begin{bmatrix}
0  & 
-i \\
i  
&
~~0
\end{bmatrix} \sin(\theta) = -i \hat{\alpha}_z \cdot  \frac{\partial}{\partial z}  + \hat{\beta}_z \sin(\theta).
\eea
See Appendix~\ref{app1} for the intermediate steps. The standard form of the two-component Dirac Hamiltonian is,  
\be
{\bf \hat{H}_D} = -i\hbar c \hat{\alpha}\cdot\frac{\partial}{\partial z}  + \hat{\beta}m c^2,
\ee
where $m$ is the mass of the particle, $c$ is the speed of light, $z$ is the spatial coordinate, matrix, $\hat{\alpha}$ and $\hat{\beta}$ are Hermitian and satisfy, $\hat{\alpha}^2 = \hat{\beta}^2 = {\mathbbm 1}$ and $\hat{\alpha}\hat{\beta} = - \hat{\beta}\hat{\alpha}$.  For a two-component Dirac Hamiltonian, $\hat{\alpha}$ and $\hat{\beta}$ are the Pauli matrices.
In Eq.\,(\ref{hamil2aa}), though the matrices $\hat{\alpha}_z$ and $\hat{\beta}_z$
are Hermitian and satisfy, $\hat{\alpha}_z^2 = \hat{\beta}_z^2 = {\mathbbm 1}$ and $\hat{\alpha}_z\hat{\beta}_z = - \hat{\beta}_z\hat{\alpha}_z$,  $\hat{\alpha}_z$ is not a Pauli matrix. Since the Hamiltonian has a rotational invariance property, by introducing rotation $\hat{R}_y(\theta/2) = e^{-i\frac{\theta}{2}\hat{\sigma}_y}$  to the Eq.\,(\ref{hamil2aa}), $\hat{\alpha}_z$ changes to the Pauli matrix,
\bea
\label{dirac1}
\hat{H}^{\prime}_{\sigma_{z}}(\theta) = \hat{R}^{\dagger}_y\left (\frac{\theta}{2} \right ) \hat{H}_{\sigma_{z}}(\theta)\hat{R}_y \left ( \frac{\theta}{2} \right ) = -i \hat{\sigma}_z \cdot \frac{\partial}{\partial z} + \hat{\sigma}_y \sin(\theta).
\eea
The Hamiltonian given by Eq.\,(\ref{dirac1}) is identical to the two-component Dirac Hamiltonian, the mass equivalent term, $m = \sin(\theta)$ and $c = 1$.

\noindent
{\bf Basis states of $\hat{\sigma}_x$ as position translation state~:~}
For the basis states,
$ |+\rangle_{\sigma_{x}} = \frac{1}{\sqrt 2} \begin{bmatrix} 1 \\ 1 \end{bmatrix} ~;~ |-\rangle_{\sigma_{x}}= \frac{1}{\sqrt 2} \begin{bmatrix} ~1 \\ -1 \end{bmatrix}$
of the Pauli operator $\hat{\sigma}_{x} = \begin{bmatrix} 0 & 1\\ 1 & 0 \end{bmatrix}$ as the position translation states, we will choose $|x\rangle$ 
as the position space where $x \in  {\mathbbm  I}$. The quantum coin operation for the walk will be, 
\bea
\label{coinP}
\hat{B}_{\sigma_{x}}(\theta)& = &\cos (\theta) |+ \rangle_{\sigma_{x}} \langle + |
 + \sin (\theta) |+ \rangle_{\sigma_{x}} \langle -|  - \sin (\theta) |- \rangle_{\sigma_{x}} \langle + |
+ \cos (\theta) |- \rangle_{\sigma_{x}} \langle - |,
\eea
and the shift operator will be,
\bea
\label{shiftsigma1}
\hat{S}_{\sigma_x} \equiv \sum_{x}  [ |+ \rangle_{\sigma_{1}}\langle
+|\otimes|x-1\rangle\langle    x| 
+    | - \rangle_{\sigma_{1}}\langle  
-|\otimes |x+1\rangle\langle  x|].
\eea
The effective operator for one complete step of the walk will be, 
$\hat{W}_{\sigma_x} (\theta) = \hat{S}_{\sigma_x} [ \hat{B}_{\sigma_{x}}(\theta) \otimes  {\mathbbm 1}]$.
When the initial state of the two-component wavefunction of the particle at the origin is given by $|\Psi_{in} \rangle =   \begin{bmatrix} 
  \psi^{\downarrow}(0, 0) \\
  \psi^{\uparrow}(0, 0) \\
  \end{bmatrix}$, the state after $t$ steps of the walk will be,
  \bea
|\Psi (t) \rangle = [\hat{W}_{\sigma_x} (\theta)]^t |\Psi_{in}\rangle =   \sum_x \begin{bmatrix} 
  \psi^{\downarrow}(x, t) \\
  \psi^{\uparrow}(x, t) \\
  \end{bmatrix},
\eea
where $\hat{W}_{\sigma_x} (\theta)$ when written in the basis formed by the eigenvectors of $\hat{\sigma}_{z}$ is, 
\begin{align}
\hat{W}_{\sigma_x} (\theta) \equiv \frac{e^{-i\hat{P}_x}}{2} \begin{bmatrix}
   \cos(\theta)+ \sin(\theta)   &  &   \cos(\theta)-\sin(\theta)\\ 
  \cos(\theta)+\sin(\theta) &  & \cos(\theta)-\sin(\theta) 
\end{bmatrix} + 
  \frac{e^{+i\hat{P}_x}}{2} \begin{bmatrix}
 \cos(\theta)-\sin(\theta)  &  & -\cos(\theta)-\sin(\theta)\\ 
 -\cos(\theta)+\sin(\theta) & & \cos(\theta)+\sin(\theta)
\end{bmatrix} \equiv e^{-i\hat{H}_{\sigma_x}(\theta)}.
 \label{eq:combop33}
\end{align}
Here $\hat{P}_x$ is the position displacement operator whose action on state $\psi^{\downarrow (\uparrow)} (x, t)$ is local such that $e^{\pm i \hat{P}_x} \psi^{\downarrow (\uparrow)} (x, t) = \psi^{\downarrow (\uparrow)} (x \pm 1, t)$. By taking the logarithm of Eq.\,(\ref{eq:combop33}) and simplifying we obtain the effective Hamiltonian for each step of the walk on the discrete position space, 
\bea
\label{hamil2b}
\hat{H}_{\sigma_{x}}(\theta)= \frac{\hat{\omega}_x}{\sin(\hat{\omega}_x)}
\begin{bmatrix}
\cos(\theta)\sin(\hat{P}_x)  - i \sin(\theta)\cos(\hat{P}_x) \,  & 
\sin(\theta) \sin(\hat{P}_x)\, \\
-\sin(\theta) \sin(\hat{P}_x) \, 
&
\cos(\theta)\sin(\hat{P}_x)  + i \sin(\theta) \cos(\hat{P}_x)
\end{bmatrix} \cdot \hat{\sigma}_x. 
\eea
Here $\hat{\omega}_x = \cos^{-1}\Big (\cos(\theta)\cos(\hat{P}_x)\Big)$ and the  differential operator form of the Hamiltonian [Eq.\,(\ref{hamil2b})] is,
\bea
\label{hamil2ba}
\hat{H}_{\sigma_{x}}(\theta)=  -i \begin{bmatrix}
\sin(\theta)   & 
~~\cos(\theta) \\
 \cos(\theta) 
&
- \sin(\theta) 
\end{bmatrix} \cdot  \frac{\partial}{\partial x} + \begin{bmatrix}
~~0  & 
~i \\
-i  
&
~0
\end{bmatrix} \sin(\theta) = -i \hat{\alpha}_x  \cdot \frac{\partial}{\partial x}  + \hat{\beta}_x \sin(\theta).
\eea
The matrices $\hat{\alpha}_x$ and $\hat{\beta}_x$ are Hermitian and satisfy,  $\hat{\alpha}^2_x = \hat{\beta}^2_x = {\mathbbm 1}$ and $\hat{\alpha}_x\hat{\beta}_x = - \hat{\beta}_x\hat{\alpha}_x$.  We can obtain the Dirac form of the Hamiltonian by introducing rotation $\hat{R}_y(\theta/2) = e^{-i\frac{\theta}{2}\hat{\sigma}_y}$  to the Eq.\,(\ref{hamil2ba}),
\bea
\label{dirac2}
\hat{H}^{\prime}_{\sigma_{x}}(\theta) = \hat{R}_y\left (\frac{\theta}{2} \right ) \hat{H}_{\sigma_{x}}(\theta)\hat{R}^{\dagger}_y \left ( \frac{\theta}{2} \right ) = -i \hat{\sigma}_x \cdot \frac{\partial}{\partial x} - \hat{\sigma}_y \sin(\theta).
\eea
The Hamiltonian given by Eq.\,(\ref{dirac2}) is identical to the two-component Dirac Hamiltonian, with the mass equivalent term, $m = -\sin(\theta)$ and the velocity equivalent term, $c = 1$.

\vskip 0.2in
\noindent
{\bf Basis states of $\hat{\sigma}_y$ as position translation state~:}
For the basis states
$|+\rangle_{\sigma_{y}} = \frac{1}{\sqrt 2} \begin{bmatrix} 1 \\ i \end{bmatrix} ~;~ |-\rangle_{\sigma_{2}}= \frac{1}{\sqrt 2} \begin{bmatrix} ~~1 \\ -i \end{bmatrix}$
of Pauli operators $\hat{\sigma}_{y} = \begin{bmatrix} 0  & -i\\ i & ~~0 \end{bmatrix}$ as the translation states, we will choose $|y\rangle$ where $y \in  {\mathbbm  I}$ as the position space. The coin operation for the walk will be,
\bea
\label{coinP}
\hat{B}_{\sigma_{y}}(\theta)& = &\cos (\theta) |+ \rangle_{\sigma_{y}} \langle + |
 + \sin (\theta) |+ \rangle_{\sigma_{y}} \langle -|    - \sin (\theta) |- \rangle_{\sigma_{y}} \langle + |
+ \cos (\theta) |- \rangle_{\sigma_{y}} \langle - |,
\eea
and the shift operator will be,
\bea
\label{shiftsigma1}
\hat{S}_{\sigma_y} \equiv \sum_{y}  [ |+ \rangle_{\sigma_{y}}\langle
+|\otimes|y-1\rangle\langle    y| 
+    | - \rangle_{\sigma_{y}}\langle  
-|\otimes |y+1\rangle\langle  y|].
\eea
The effective operator for each step of the walk will be, 
$\hat{W}_{\sigma_y} (\theta) = \hat{S}_{\sigma_y} [ \hat{B}_{\sigma_{y}}(\theta) \otimes  {\mathbbm 1}]$.
When the initial state of the particle at the origin, $|\Psi_{in}\rangle =   \begin{bmatrix} 
  \psi^{\downarrow}(0, 0) \\
  \psi^{\uparrow}(0, 0) \\
  \end{bmatrix}$, the state at time $t$ will be,
  \bea
|\Psi (t) \rangle = [\hat{W}_{\sigma_y} (\theta)]^t |\Psi_{in} \rangle =   \sum_y \begin{bmatrix} 
  \psi^{\downarrow}(y, t) \\
  \psi^{\uparrow}(y, t) \\
  \end{bmatrix},
\eea
where $\hat{W}_{\sigma_y} (\theta)$ when written in the basis formed by the eigenvectors of $\hat{\sigma}_{z}$ is, 
\begin{align}
 \hat{W}_{\sigma_y} (\theta)  =\frac{e^{-i\hat{P}_y}}{2}\begin{bmatrix}
   \cos(\theta)+ \sin(\theta)         &    -i\cos(\theta)+i\sin(\theta)
  \\ i\cos(\theta)+i\sin(\theta)  &  \cos(\theta)-\sin(\theta) 
\end{bmatrix} 
  +\frac{e^{i\hat{P}_y}}{2} \begin{bmatrix}
  \cos(\theta)-\sin(\theta)           &  i\cos(\theta)+i\sin(\theta)
  \\ -i\cos(\theta)+\sin(\theta) & \cos(\theta)+\sin(\theta) 
\end{bmatrix} \equiv e^{-i\hat{H}_{\sigma_y}(\theta)} 
\label{eq:combop4}.
\end{align}
Here $\hat{P}_y$ is the position displacement operator whose action on state $\psi^{\downarrow (\uparrow)} (y, t)$ is local such that $e^{\pm i \hat{P}_y} \psi^{\downarrow (\uparrow)} (y, t) = \psi^{\downarrow (\uparrow)} (y \pm 1, t)$. By taking logarithm of the Eq.\,(\ref{eq:combop4}) and simplifying we obtain the the effective Hamiltonian form for each step of the walk on the discrete position space,  
\bea
\label{hamil2c}
\hat{H}_{\sigma_{y}}(\theta) =  \frac{\hat{\omega}_y}{\sin(\hat{\omega}_y)}
\begin{bmatrix}
\cos(\theta)\sin(\hat{P}_y)-i\sin(\theta) \cos(\hat{P}_y) & 
-i\sin(\theta) \sin(\hat{P}_y)\\
-i\sin(\theta) \sin(\hat{P}_y) 
&
\cos(\theta)\sin(\hat{P}_y)+i\sin(\theta)\cos(\hat{P}_y)  
\end{bmatrix} \cdot \hat{\sigma}_y.
\eea
Here $\hat{\omega}_y = \cos^{-1}\Big (\cos(\theta)\cos(\hat{P}_y)\Big)$ and the differential operator form of the Hamiltonian [Eq.\,(\ref{hamil2c})] is,
\bea
\label{hamil2ca}
\hat{H}_{\sigma_{y}}(\theta)=  -i \begin{bmatrix}
~~\sin(\theta)   & 
-i\cos(\theta) \\
~i \cos(\theta) 
&
- \sin(\theta) 
\end{bmatrix}  \cdot  \frac{\partial}{\partial y} + \begin{bmatrix}
0  & 
~1 \\
1  
&
~0
\end{bmatrix} \sin(\theta) = -i \hat{\alpha}_y  \cdot \frac{\partial}{\partial y}  + \hat{\beta}_y \sin(\theta).
\eea
The matrices $\hat{\alpha}_y$ and $\hat{\beta}_y$ are Hermitian and satisfy  $\hat{\alpha}^2_y = \hat{\beta}^2_y = {\mathbbm 1}$ and $\hat{\alpha}_y\hat{\beta}_y = - \hat{\beta}_y\hat{\alpha}_y$.  

We obtain the Dirac form of the Hamiltonian by introducing rotation $\hat{R}_x(\theta/2) = e^{-i\frac{\theta}{2}\hat{\sigma}_x}$  to the Eq.\,(\ref{hamil2ca}),
\bea
\label{dirac3}
\hat{H}^{\prime}_{\sigma_{y}}(\theta) = \hat{R}^{\dagger}_x\left (\frac{\theta}{2} \right ) \hat{H}_{\sigma_{y}}(\theta)\hat{R}_x \left ( \frac{\theta}{2} \right ) = -i \hat{\sigma}_y \cdot \frac{\partial}{\partial y} + \hat{\sigma}_x \sin(\theta).
\eea
The Hamiltonian given by the Eq.\,(\ref{dirac3}) is identical to the two-component Dirac Hamiltonian with the mass equivalent term, $m = \sin(\theta)$ and velocity term, $c = 1$. When the coin operation parameter $\theta = 0$, all the three forms of Hamiltonian obtained for DQW evolution will be identical to the two-component Dirac Hamiltonian for the massless particle.

\section{The Hamiltonian for a walk on two- and three- dimensional lattice}
\begin{figure}[ht]
\bc
\includegraphics[width=10.0cm]{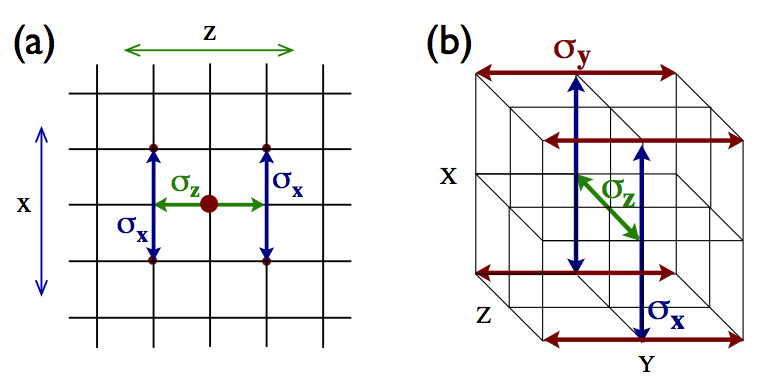}
\caption{{\bf Schematic for the DTQW showing the direction of evolution.~~}(a) Square lattice with two direction for propagation along each quantization axis, $X$ and $Z$. (b) Cubic lattice with two direction of propagation along each quantization axis, $X$, $Y$ and $Z$. Evolution along each quantization axis, $X$, $Y$ and $Z$ are defined using basis states of the Pauli operators $\hat{\sigma}_{x}$, $\hat{\sigma}_{y}$ and $\hat{\sigma}_{z}$ as position translation states.\label{fig:1}}
\ec 
\end{figure}
\noindent
{\bf Square Lattice :} 
A simple 2D lattice structure is a square lattice with four direction for propagation at each position and the evolution can be quantized along the axis, $X$ and $Z$ [Fig.\,\ref{fig:1}(a)]. One of the well studied schemes for 2D DQW is the Grover walk on a four-state particle with the basis states as $|0\rangle$, $|1\rangle$, $|2\rangle$, and $|3\rangle$ \cite{TFM03, WKK08}. Each step of Grover walk on a 2D is realized using the Grover diffusion operator, 
\bea
\hat{G} = \frac{1}{2}\begin{bmatrix}-1 & ~~1 & ~~1 & ~~1  \\
  ~~1 & -1 & ~~1 & ~~1  \\
  ~~1 & ~~1 & -1 & ~~1  \\
  ~~1 & ~~1 & ~~1 & -1
\end{bmatrix},
\eea
as coin operation followed by a shift operator, 
\bea
  \hat{S}^G \equiv \sum_{x, z} \Big[|0 \rangle\langle 0|\otimes|
                          x-1, z-1\rangle\langle x, z|   
                        +     |1 \rangle\langle 1 |\otimes |
                          x-1, z+1\rangle\langle x, z|  \nonumber \\
                        +     | 2 \rangle\langle 2 |\otimes|
                          x+1, z-1\rangle\langle   x, z| 
                        +     |3 \rangle\langle 3|\otimes |
                          x+1, z+1\rangle\langle   x, z| \Big]
                          \eea
                           on a particle in a specific initial state, $|\Psi_{\rm in}^{4s}\rangle=\frac{1}{2}[|0 \rangle -|1 \rangle -|2 \rangle +|3 \rangle]$. 
The state after $t$ steps of the Grover walk, [$\hat{S}^G(\hat{G} \otimes  {\mathbbm 1})]^t$,
\bea 
|\Psi^G_t\rangle = \sum_{x =-t}^t \sum_{z =-t}^t \Big [ \beta^{(1)}_{(x, z, t)}|0
  \rangle  + \beta^{(2)}_{(x, z, t)}|1 \rangle  
+  \beta^{(3)}_{(x, z, t)}|2 \rangle 
 + \beta^{(4)}_{(x, z, t)}|3 \rangle \Big ] 
\otimes|x,  z\rangle.
\label{eq:lr2}
\eea
Here $\beta(x, y,t)$'s  are given by the
quadrupled iterative relation coupling the $X$ and $Z$ axis
\begin{widetext}
\begin{subequations}
  \label{eq:4s_iter}
  \begin{eqnarray}
   \beta^{(1)}(x, z, t) &= \frac{1}{2} \Big[-\beta^{(1)}_{(x+1, z+1,t-1)} + \beta^{(2)}_{(x+1, z+1,t-1)} 
     &+ \beta^{(3)}_{(x+1, z+1,t-1)} +  \beta^{(4)}_{(x+1, z+1,t-1)}\Big ], \\
   \beta^{(2)}(x, z, t) &= \frac{1}{2} \Big[\beta^{(1)}_{(x+1, z-1,t-1)} -  \beta^{(2)}_{(x+1, z-1,t-1)}  
    &+ \beta^{(3)}_{(x-1, z-1,t-1)} +  \beta^{(4)}_{(x+1, z-1,t-1)} \Big], \\
    \beta^{(3)}(x, z, t) &= \frac{1}{2} \Big[\beta^{(1)}_{(x-1, z+1,t-1)} + \beta^{(2)}_{(x-1, z+1,t-1)} 
    & - \beta^{(3)}_{(x-1, z+1,t-1)}  + \beta^{(4)}_{(x-1, z+1,t-1)}\Big ],\\
    \beta^{(4)}(x, z, t) &= \frac{1}{2} \Big [\beta^{(1)}_{(x-1, z-1,t-1)} +  \beta^{(2)}_{(x-1, z-1,t-1)} 
    & + \beta^{(3)}_{(x-1, z-1,t-1)} -  \beta^{(4)}_{(x-1, z-1,t-1)} \Big ].
  \end{eqnarray}
\end{subequations}
           \end{widetext}
\par
DQW on a square lattice using two-state particle can also be realized by quantizing the evolution using different Pauli basis states as translation state for each axis in the lattice structure.  That is, each step of walk on a square lattice comprise of evolution in $Z-$ axis with Pauli basis of $\hat{\sigma}_z$ operator as translational state followed by the evolution in the $X-$axis with Pauli basis of $\hat{\sigma}_x$ operator as translational state,  
\bea
\label{w2d}
\hat{W}^{sq}(\theta) = \hat{W}_{\sigma_x}^{sq} (\theta) \hat{W}_{\sigma_z}^{sq} (\theta).
\eea
Here 
$\hat{W}_{\sigma_\alpha}^{sq} (\theta)= \hat{S}_{\sigma_\alpha}^{sq} [ \hat{B}_{\sigma_{\alpha}}(\theta) \otimes  {\mathbbm 1}_X\otimes  {\mathbbm 1}_Z]$ with $\alpha = x, z$  and the shift operators for $Z$- and $X$- axis are, 
\begin{subequations}
\begin{eqnarray}
\label{shiftSq}
\hat{S}_{\sigma_z}^{sq} \equiv \sum_{x, z}  [ |+ \rangle_{\sigma_{z}}\langle
+|\otimes|x, z-1\rangle\langle   x, z|  
+    | - \rangle_{\sigma_{z}}\langle
-|\otimes |x, z+1\rangle\langle x, z|],  \\
\hat{S}_{\sigma_x}^{sq} \equiv \sum_{x, z}  [ |+ \rangle_{\sigma_{x}}\langle
+|\otimes|x-1, z\rangle\langle   x, z|  
+    | - \rangle_{\sigma_{x}}\langle  
-|\otimes |x+1, z\rangle\langle x, z|].
\end{eqnarray}
\end{subequations}
The choice of a particular Pauli basis for particular axis is purely conventional.
If the initial state of the particle on a square lattice at origin $(x,z) = (0, 0)$ is $|\Psi_{in}\rangle=\frac{1}{\sqrt{2}}[|\downarrow\rangle + i |\uparrow \rangle] \otimes |0, 0\rangle$, the state after $t$ steps,
\bea 
|\Psi_t\rangle =  (\hat{W}^{sq}(\theta))^t |\Psi_{in}\rangle = \sum_{x =-t}^t \sum_{z =-t}^t \Big [ \alpha^{(x)}_{(x, z, t)}|\downarrow
  \rangle  + \alpha^{(z)}_{(x, z, t)}|\uparrow \rangle \Big ] 
\otimes|x,  z\rangle.
\label{eq:lr2}
\eea
When $\theta =0$,  $\alpha^{(1)}_{(x, z,t)}$ and $\alpha^{(2)}_{(x, z, t)}$ are given by the
coupled iterative relations
\begin{subequations}
\label{eq:iter}
\begin{eqnarray}
\alpha^{(1)}_{(x, z,t)} &=& \frac{1}{2}\Bigg [  \alpha^{(1)}_{(x+1, z+1,t-1)}  + \alpha^{(1)}_{(x+1, z-1,t-1)}  + \alpha^{(2)}_{(x-1, z+1,t-1)} -  \alpha^{(2)}_{(x-1, z-1,t-1)}\Bigg ],  \\
\alpha^{(2)}_{(x, z,t)} &=& \frac{1}{2}\Bigg [  \alpha^{(1)}_{(x+1, z+1,t-1)} -  \alpha^{(1)}_{(x+1, z-1,t-1)} + \alpha^{(2)}_{(x-1, z+1,t-1)} +  \alpha^{(2)}_{(x-1, z-1,t-1)}\Bigg ].
\end{eqnarray}
\end{subequations}
From Eqs.\,(\ref{eq:iter}) and Eqs.\,(\ref{eq:4s_iter}) we can note that for both, two-state walk and the Grover walk, the amplitude at any position $(x, z)$ for a given time $t$ is dependent on the amplitude at the four diagonally opposite sites at time $t-1$. 
\par
\begin{figure}[ht]
\bc
\includegraphics[width=150mm]{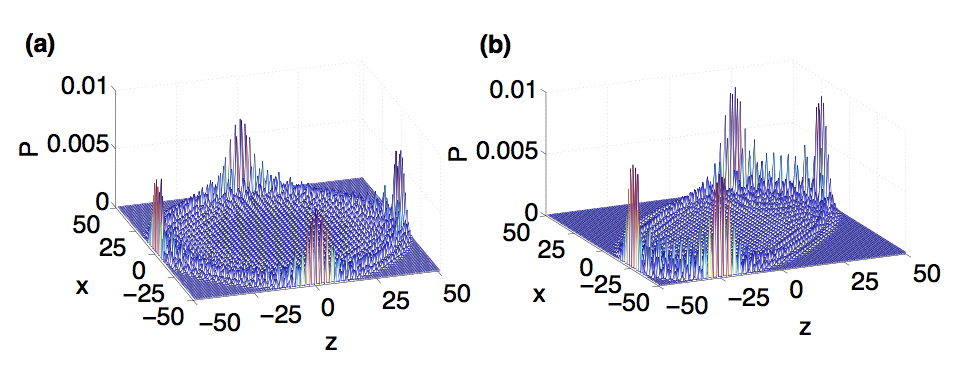}
\caption{{\bf Probability distribution of DQW on square lattice without and with coin operation.} Probability distribution after a DQW on a two-state particle at the origin $(x, z) = (0,0)$ with initial state, $|\Psi_{\rm in} \rangle = \frac{1}{\sqrt 2}[|\downarrow \rangle + i|\uparrow \rangle]\otimes|0, 0\rangle$ on a square lattice using basis state of $\hat{\sigma}_x$ and  $\hat{\sigma}_z$ as position translation state for evolution along $X$ and $Z$ axis, respectively. The distribution is after 50 step of the walk. (a) The distribution is after the evolution without using a coin operation ($\theta =0$) and same distribution is obtained for Grover walk (with the specific initial state). (b) The distribution is after the evolution with the coin operation $\theta = \pi/12$ in both the axis.\label{fig:2} }
\ec
\end{figure}
In Fig.\,\ref{fig:2}(a), the probability distribution of the 50 step DQW on a square lattice using the Pauli basis scheme without the external coin operation ($\theta = 0$) is shown. We notice that the probability distribution is identical to the distribution obtained for the Grover walk on a four-state particle\,\cite{TFM03} and for the alternative walk on a two-state particle with initial state, $\frac{1}{\sqrt{2}}[|\downarrow \rangle + i |\uparrow \rangle]$ using Hadamard operator as the coin operation for each dimension\,\cite{FGB11, CB13}. The main reason for this is the fact that the basis states of $\hat{\sigma}_z$ is also a superposition of a basis states of the $\hat{\sigma}_x$ which inherently introduces an effect of quantum coin operation. Therefore, the operator $[\hat{W}^{sq}_{\sigma_{x}}(0)\hat{W}^{sq}_{\sigma_{z}}(0)]^t$ continues to evolve the particle in superposition of position space bringing in intricate features into the interference effect which is showing up in the probability distribution. Unlike the Grover walk which is very specific to the initial state and the coin operation, probability distribution with the two-state walk using the Pauli basis can be controlled by introducing the coin operation ($\theta \neq 0$) and/or using different initial state of the particle. In Fig.\,\ref{fig:2}(b) the probability distribution after 50 step DQW with an external coin operation with $\theta = \pi/12$ is show to squeeze the distribution towards the diagonal of the square lattice and from this we can infer that the coin operation can be effectively used to control the dynamics and the probability distribution of the walk. \\
\vskip 0.1in
\noindent
{\it Two-component Hamiltonian for walk on a square lattice~-~}
The unitary operator  for each step of a two-state DQW on a square lattice is  $\hat{W}^{sq}(\theta) = \hat{W}_{\sigma_x}^{sq} (\theta) \hat{W}_{\sigma_z}^{sq} (\theta)$. For the initial state of the particle at origin $(x, z) = (0, 0)$,  $|\Psi^{sq}_{in} \rangle =   \begin{bmatrix} 
  \psi^{\downarrow}(0, 0, 0) \\
  \psi^{\uparrow}(0, 0, 0) \\
  \end{bmatrix}$, the state at any time $t$ will be,
  \bea
|\Psi^{sq}(t) \rangle = [\hat{W}_{\sigma_x}(\theta)\hat{W}_{\sigma_z} (\theta)]^t |\Psi^{sq}_{in} \rangle =   \sum_{x, z} \begin{bmatrix} 
  \psi^{\downarrow}(x, z, t) \\
  \psi^{\uparrow}(x, z, t) \\
  \end{bmatrix},
\eea
where $\hat{W}_{\sigma_z}(\theta)$ and $\hat{W}_{\sigma_x}(\theta)$ are given by Eqs.\,(\ref{paulizshif}) and (\ref{eq:combop33}), respectively. The effective Hamiltonian for each step of the walk on a square lattice is the sum of the Hamiltonian for evolution in each dimension. In differential operator form it will be,
\bea
\hat{H}^{sq}(\theta) = \hat{H}_{\sigma_x}(\theta) + \hat{H}_{\sigma_z}(\theta)   = - i \Big ( \hat{\alpha}_x \cdot \frac{\partial}{\partial x} +  \hat{\alpha}_z \cdot \frac{\partial}{\partial z} \Big ) + (\hat{\beta}_x  +  \hat{\beta}_z)\sin(\theta).
\eea
Here the matrix $\hat{\alpha}_z$, $\hat{\alpha}_x$, $\hat{\beta}_z$, $\hat{\beta}_x$ are same as that in Eqs.\,(\ref{hamil2aa}) and \,(\ref{hamil2ba}).  As the evolution in one dimension is followed by the other, the rotational invariance property of the Hamiltonian can be used independently to the Hamiltonian for each dimension resulting in, 
\bea
\hat{H}^{{sq}^{\prime}}(\theta) = \hat{H}^{\prime}_{\sigma_x}(\theta) + \hat{H}^{\prime}_{\sigma_z}(\theta)   = - i \Big ( \hat{\sigma}_x \cdot \frac{\partial}{\partial x} +  \hat{\sigma}_z \cdot \frac{\partial}{\partial z} \Big ),
\eea
where $\hat{H}^{\prime}_{\sigma_z}(\theta)$ and $\hat{H}^{\prime}_{\sigma_x}(\theta)$ are same as Eqs.\,(\ref{dirac1}) and \,(\ref{dirac2}). As $\hat{\beta}_{x} +\hat{\beta}_{z} = 0$ this Hamiltonian is structurally identical to the two-component Dirac Hamiltonian for a massless particle on a two-dimensional space. In the Hamiltonian obtained from 1D DQW evolution operator, the mass equivalent term becomes zero only when $\theta=0$,  but from the 2D evolution operator we obtain an Hamiltonian which resembles the massless Dirac Hamiltonian even for non-zero value of $\theta$.  However,  by choosing basis state of $\hat{\sigma}_y$ operator as the position translation state for one of the dimension and basis state of $\hat{\sigma}_z$ or $\hat{\sigma}_x$ operator for an other dimension we can obtain an Hamiltonian with non-zero mass equivalent term. 

\begin{figure}[ht]
\bc
\includegraphics[width=14.6cm]{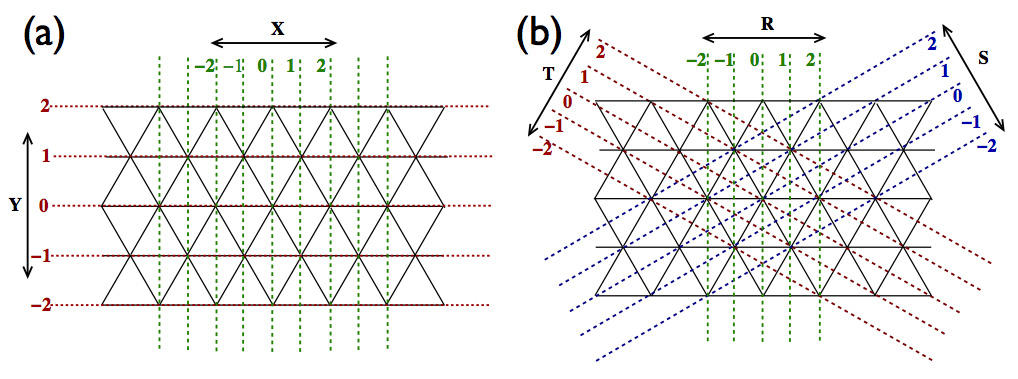}
\caption{{\bf Schematic showing the labelling and evolution axis for triangular lattice.}~~Triangular lattice structure with (a) labeling of lattice positions in the spatial coordinates and (b) the three axis, $R$, $S$, and $T$ used as quantization axis for the evolution using different Pauli basis states as translation states. \label{fig:3}}
\ec
\end{figure}

\vskip 0.2in
\noindent
{\bf Cubic Lattice~:~}
A simple 3D lattice is a cubic lattice with six direction for propagation at each position and the evolution can be quantized along the axis, $X$, $Y$, and $Z$ [Fig.\,\ref{fig:1}(b)]. By evolving the two-state particle in one axis followed by the evolution in the other two axis using different Pauli basis as position translation state, a two-state walk on a cubic lattice can be realized. That is,
\bea
\label{w3d}
\hat{W}^{cub}(\theta) = \hat{W}_{\sigma_y}^{cub} (\theta)\hat{W}_{\sigma_x}^{cub} (\theta) \hat{W}_{\sigma_z}^{cub} (\theta).
\eea
Here $\hat{W}_{\sigma_j}^{cub} (\theta)=\hat{S}_{\sigma_j}^{cub} [ \hat{B}_{\sigma_{j}}(\theta) \otimes  {\mathbbm 1}_X\otimes  {\mathbbm 1}_Y \otimes  {\mathbbm 1}_Z]$ with $j  =\{ y, x, z \}$ where the shift operator for evolution along the $Z$, $X$, and $Y$ axis will be
\begin{subequations}
\begin{eqnarray}
\hat{S}_{\sigma_z}^{cub} \equiv \sum_{x,y,z}  \Big [ |+ \rangle_{\sigma_{z}}\langle
+|\otimes|x, y,z-1\rangle\langle   x, y, z|  
+    | - \rangle_{\sigma_{z}}\langle
-|\otimes |x,y, z+1\rangle\langle x, y, z| \Big ],  \\
\hat{S}_{\sigma_x}^{cub} \equiv \sum_{x, y, z}  \Big [ |+ \rangle_{\sigma_{x}}\langle
+|\otimes|x-1, y, z\rangle\langle   x,y, z|  
+    | - \rangle_{\sigma_{x}}\langle  
-|\otimes |x+1,y, z\rangle\langle x,y, z| \Big ], \\
\hat{S}_{\sigma_y}^{cub} \equiv \sum_{x, y, z}  \Big [ |+ \rangle_{\sigma_{y}}\langle
+|\otimes|x, y-1, z\rangle\langle   x,y, z|  
+    | - \rangle_{\sigma_{y}}\langle  
-|\otimes |x,y+1, z\rangle\langle x,y, z| \Big ].
\end{eqnarray}
\end{subequations}
For the initial state of the particle at origin $(x, y, z) = (0, 0, 0)$,  $|\Psi^{cub}_{in} \rangle =   \begin{bmatrix} 
  \psi^{\downarrow}(0, 0, 0, 0) \\
  \psi^{\uparrow}(0, 0, 0, 0) \\
  \end{bmatrix}$, the state at any time $t$,
  \bea
|\Psi^{cub}(t) \rangle = [\hat{W}_{\sigma_y}(\theta)\hat{W}_{\sigma_x}(\theta)\hat{W}_{\sigma_z} (\theta)]^t |\Psi^{cub}_{in} \rangle =   \sum_{x, y, z} \begin{bmatrix} 
  \psi^{\downarrow}(x, y, z, t) \\
  \psi^{\uparrow}(x, y, z, t) \\
  \end{bmatrix},
\eea
where $\hat{W}_{\sigma_z}(\theta)$, $\hat{W}_{\sigma_x}(\theta)$  and $\hat{W}_{\sigma_y}(\theta)$ are given by Eqs.\,(\ref{paulizshif}),  (\ref{eq:combop33}) and (\ref{eq:combop4}), respectively. Therefore the effective Hamiltonian for each step of the walk on a cubic lattice is,
\bea
\hat{H}^{cub}(\theta) = \hat{H}_{\sigma_y}(\theta)  + \hat{H}_{\sigma_x}(\theta) + \hat{H}_{\sigma_z}(\theta)   = -i \Big ( \hat{\alpha}_y \cdot  \frac{\partial}{\partial y} + \hat{\alpha}_x \cdot \frac{\partial}{\partial x} +  \hat{\alpha}_z \cdot \frac{\partial}{\partial z} \Big ) + \hat{\beta}_y \sin(\theta).
\eea
Here the matrix $\hat{\alpha}_z$, $\hat{\alpha}_x$, $\hat{\alpha}_y$, $\hat{\beta}_y$ are same as that in Eqs.\,(\ref{hamil2aa}),\,(\ref{hamil2ba}), and (\ref{hamil2ca}) 
(where $\hat{\beta}_{x} +\hat{\beta}_{z} = 0$). As the evolution in one dimension is followed by the other, the rotational invariance property of the Hamiltonian can be used independently to the Hamiltonian for each dimension on a cubic lattice resulting in, 
\bea
\hat{H}^{{cub}^{\prime}}(\theta) = \hat{H}^{\prime}_{\sigma_x}(\theta)  + \hat{H}^{\prime}_{\sigma_y}(\theta) + \hat{H}^{\prime}_{\sigma_z}(\theta)   = -i \Big ( \hat{\sigma}_x \cdot  \frac{\partial}{\partial x} + \hat{\sigma}_y \cdot \frac{\partial}{\partial y} +  \hat{\sigma}_z \cdot \frac{\partial}{\partial z} \Big ) + \hat{\beta}_y \sin(\theta),
\eea
where $\hat{H}^{\prime}_{\sigma_z}(\theta)$, $\hat{H}^{\prime}_{\sigma_x}(\theta)$, and $\hat{H}^{\prime}_{\sigma_y}(\theta)$ are same as Eqs.\,(\ref{dirac1}), \,(\ref{dirac2})
 and \,(\ref{dirac3}). The Hamiltonian from the evolution operator for DQW on a cubic lattice is structurally identical to the two-component Dirac Hamiltonian for a particle on a three-dimensional space. Due to the fact that the basis state of one of the Pauli operator is also a superposition of the basis states of the other Pauli operator, the effect of the quantum coin operation is inherently introduced. Therefore, even in absence of external coin operation ($\theta=0$), the particle will continue to evolve in superposition of different location at each step bringing in the intricate interference effect into the evolution. 
\vskip 0.2in
\noindent
{\bf Triangular lattice :}
The triangular lattice is a 2D structure with six direction for propagation at each position.  As shown in Fig.\,\ref{fig:3}(a), each position can be labeled using indices $(x, y)$ of the $X$ and $Y$ spatial dimensions and the evolution can be quantized along three axis $R$, $S$, and $T$ as shown in Fig.\,\ref{fig:3}(b) using the basis states $|+\rangle_{\sigma_{j}}$ and $|-\rangle_{\sigma_{j}}$ of the Pauli operators $\hat{\sigma}_{j}$ where $j  =\{x, y, z\}$ as position translation states. Because of the three quantization axis for evolution on a 2D lattice, the evolution along one of the spatial dimension will also result in the evolution along the other spatial dimension. In our scheme for DQW on a triangular lattice we will define the evolution such that, along the quantization axis  $R$, the shift is only along the $X$ spatial dimension and 
along the quantization axis $S$ and $T$,  the shift in $X$ spatial dimension will also result in shift along the $Y$ spatial dimension and vice versa. 
\par

\begin{figure}[ht]
\bc
\includegraphics[width=14.6cm]{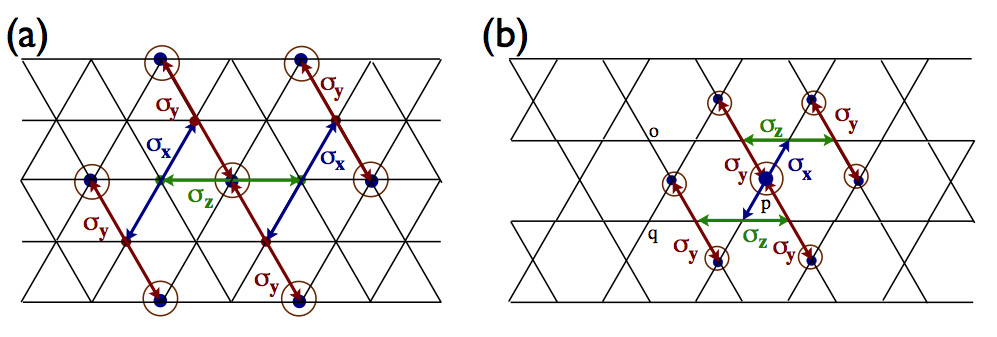}
\caption{ {\bf Schematic for evolution of DQW on triangular and kagome lattice.} (a) The triangular lattice, starting from the middle, the arrow marks indicates  the shift in position space during one step of DQW evolution (evolution in $Z$ axis followed by the evolution in the $Y$ and $X$ axis). (b) Kagome lattice structure with two axis of propagation at each lattice site. From lattice sites {\bf o}, {\bf p}, and {\bf q}, we can see that they are associated with different combination of quantization axis.  Starting from position {\bf p}, the arrow marks show the shift in position space during one step of DQW evolution. The final positions are encircled.\label{fig:4}}
\ec
\end{figure}
In Fig.\,\ref{fig:4}(a), the schematic for the evolution in one quantization axis followed by the other to implement each step of the walk is show. The encircled positions are the points in position space to which the quantum states evolve in superposition during one step of the walk. The choice of the order of the basis states as translation state to evolve the particle and the labelling of the position space is purely conventional. For the scheme used in this report, each step of the walk on a triangular lattice composes of the operation,
\bea
\hat{W}^{tri}(\theta) = \hat{W}^{tri}_{\sigma_y}(\theta)\hat{W}^{tri}_{\sigma_x}(\theta)\hat{W}^{tri}_{\sigma_z}(\theta).
\eea
Here $\hat{W}_{\sigma_j}^{tri} (\theta)=\hat{S}_{\sigma_j}^{tri} [ \hat{B}_{\sigma_{j}}(\theta) \otimes  {\mathbbm 1}_X\otimes  {\mathbbm 1}_Y \otimes  {\mathbbm 1}_Z]$ with $j  =\{ y, x, z \}$ where the shift operator for evolution along the quantization axis $R$, $S$, and $T$ are,
\begin{align}
\hat{S}^{tri}_{\sigma_z} &\equiv \sum_{x, y} \Big [ |+ \rangle_{\sigma_{z}}\langle
+| \otimes|x+2, y\rangle\langle   x, y|  
+    |- \rangle_{\sigma_{z}}\langle -| \otimes |x-2, y\rangle\langle x, y| \Big ], \\
\hat{S}^{tri}_{\sigma_x} &\equiv \sum_{x, y} \Big [|+ \rangle_{\sigma_{x}}\langle
+| \otimes|x+1, y+1 \rangle\langle   x, y|  
+   |- \rangle_{\sigma_{x}}\langle
-| \otimes |x-1, y-1 \rangle\langle x, y| \Big ], \\
\hat{S}^{tri}_{\sigma_y} &\equiv \sum_{x, y} \Big [ |+ \rangle_{\sigma_{y}}\langle
+| \otimes|x+1, y-1\rangle\langle   x, y|  
+    |- \rangle_{\sigma_{y}}\langle
-| \otimes |x-1, y+1\rangle\langle x, y| \Big ].
\end{align}

For the initial state of the particle at origin $(x, y) = (0, 0)$,  $|\Psi^{tri}_{in} \rangle =   \begin{bmatrix} 
  \psi^{\downarrow}(0, 0, 0) \\
  \psi^{\uparrow}(0, 0, 0) \\
  \end{bmatrix}$, the state at any time $t$,
  \bea
|\Psi^{tri}(t) \rangle = [\hat{W}^{tri}_{\sigma_y}(\theta)\hat{W}^{tri}_{\sigma_x}(\theta)\hat{W}^{tri}_{\sigma_z} (\theta)]^t |\Psi^{tri}_{in} \rangle =   \sum_{x, y} \begin{bmatrix} 
  \psi^{\downarrow}(x, y, t) \\
  \psi^{\uparrow}(x, y, t) \\
  \end{bmatrix}.
\eea
Here $\hat{W}^{tri}_{\sigma_z}(\theta)$ is identical to Eq.\,(\ref{paulizshif}) with a replacement of $\hat{P}_z$ by $2\hat{P}_{x}$,  $\hat{W}^{tri}_{\sigma_x}(\theta)$  is identical to 
Eq.\,(\ref{eq:combop33}) with a replacement of $\hat{P}_x$ by  $\hat{P}_{x} + \hat{P}_{y}$ and $\hat{W}^{tri}_{\sigma_y}(\theta)$ is identical to Eq.\,(\ref{eq:combop4}), with a replacement of $\hat{P}_y$ by  $\hat{P}_{x} - \hat{P}_{y}$.  Due to the fact that the basis state of one of the Pauli operator is also a superposition of the basis states of the other Pauli operator, inherently introducing the effect of the quantum coin operation. Therefore, even in absence of the external coin operation ($\theta=0$) the particle  continue to evolve in superposition of different location at each step bringing in the intricate interference effect into the evolution. However, a coin operation with different $\theta$ for each axis can be extensively used for the evolution to get addition freedom to control the dynamics and obtain the desired probability distribution.
\begin{figure}[ht]
\bc
\includegraphics[width=175mm]{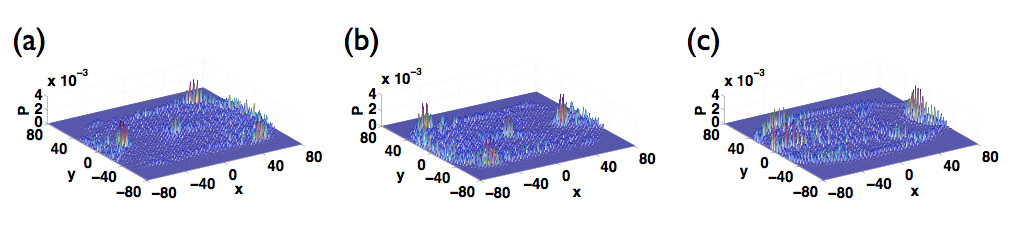}
\caption{{\bf Probability distribution after 40 step of DQW on triangular lattice.} 
(a) The initial state $|\Psi_{\rm in} \rangle = |\downarrow \rangle \otimes|0, 0\rangle$ and the walk is evolved without a coin operation. (b) The initial state $|\Psi_{\rm in}\rangle = |\uparrow \rangle \otimes |0, 0 \rangle$ and the walk is evolved without an external coin operation. (c) The initial state $|\Psi_{\rm in} \rangle = |\uparrow \rangle \otimes |0, 0\rangle$ and the walk is evolved with the coin operation along one of the quantization axis, $\hat{W}^{tri}(\theta) = \hat{W}^{tri}_{\sigma_y}(0)\hat{W}^{tri}_{\sigma_x}(\pi/4)\hat{W}^{tri}_{\sigma_z}(0)$. We can see that (a) is symmetric to (b) and an external coin operation alters the interference pattern significantly.\label{fig:5}}
\ec
\end{figure}
In Fig.\,\ref{fig:5}(a) and \,\ref{fig:5}(b), we show the probability distribution of a 40 step DQW without an external coin operation [$\hat{W}^{tri}(0) = \hat{W}^{tri}_{\sigma_{y}}(0) \hat{W}^{tri}_{\sigma_{x}}(0)\hat{W}^{tri}_{\sigma_{z}}(0)$] on a two-state particle initially in state $|\downarrow \rangle$ and $|\uparrow \rangle$, respectively.  We can see that the probability distribution in Fig.\,\ref{fig:5}(a) and Fig.\,\ref{fig:5}(b) are not symmetric distribution in position space but are symmetric to each other.  In Fig.\,\ref{fig:5}(c), we show that introducing a external coin operation along only one of the quantization axis [$\hat{W}^{tri}_{\sigma_{y}}(0)\hat{W}^{tri}_{\sigma_{x}}(\pi/4)\hat{W}^{tri}_{\sigma_{z}}(0)$], the probability distribution alters significantly. 

The effective Hamiltonian for a quantum walk on a triangular lattice is, 
\be
\hat{H}^{tri}(\theta) = \hat{H}^{tri}_{\sigma_y}(\theta) + \hat{H}^{tri}_{\sigma_x}(\theta) + \hat{H}^{tri}_{\sigma_z}(\theta),
\ee
where $\hat{H}_{\sigma_{z}}^{tri}(\theta)$, $\hat{H}_{\sigma_{x}}^{tri}(\theta)$, and $\hat{H}_{\sigma_{y}}^{tri}(\theta)$ are the Hamiltonian for evolution along quantization axis $R$, $S$, and $T$ with basis state of $\hat{\sigma}_z$, $\hat{\sigma}_x$, and $\hat{\sigma}_y$ as translation state give by,
\bea
\label{hamiltria}
\hat{H}_{\sigma_{z}}^{tri}(\theta)&=&  -i \Big ( \hat{\alpha}_z^1 \cdot  2 \frac{\partial }{\partial x}
+ \hat{\alpha}_z^2  \cdot  \frac{\partial^2}{\partial x^2} \Big)
+  \hat{\beta}_z^1
 \sin(\theta), \\
 \hat{H}_{\sigma_{x}}^{tri}(\theta)&=&  -i \Big ( \hat{\alpha}_x^1 \cdot  \frac{\partial }{\partial x} + \hat{\alpha}_x^2 \cdot   \frac{\partial }{\partial y}
+ \hat{\alpha}_x^3  \cdot  \frac{\partial^2}{\partial x \partial y} \Big)
+  \hat{\beta}_x^1 \sin(\theta), \\  
 \hat{H}_{\sigma_{y}}^{tri}(\theta)&= & -i \Big ( \hat{\alpha}_y^1 \cdot  \frac{\partial }{\partial x} + \hat{\alpha}_y^2 \cdot   \frac{\partial }{\partial y}
+ \hat{\alpha}_y^3  \cdot  \frac{\partial^2}{\partial x \partial y} \Big)
+  \hat{\beta}_y^1 \sin(\theta).
\eea
See Appendix~\ref{app2} for the matrices $\hat{\alpha}$ and $\hat{\beta}$, they are Hermitian and satisfy the conditions, $\hat{\alpha}^2 = \hat{\beta}^2 = 1$  and $\hat{\alpha}\hat{\beta} = -\hat{\beta}\hat{\alpha}$. Due to the non-orthogonal nature of the spatial axis for evolution using different basis states, we don't get an Hamiltonian which is structurally identical to the two-component Dirac Hamiltonian as we obtained for a quantum walk on a square and cubic lattice. However, the Hamiltonian with the second order differential operators can be effectively used to describes the two-state quantum walk on triangular lattice.

\vskip 0.2in
\noindent
{\bf Kagome Lattice :}
Kagome lattice structure can also be labeled the same way as the triangular lattice. The evolution operator and its Hamiltonian form in each basis  will be in the same form as presented for triangular lattice. But, unlike triangular lattice which has three quantization axis at each lattice site,  kagome lattice shown in  Fig.\,\ref{fig:4}(b) has only two quantization axis with four direction of propagation for  the walk at each lattice site. The two quantization axis at each lattice site is not the same for all lattice sites. As shown in Fig.\,\ref{fig:4}(b), lattice sites {\bf o}, {\bf p}, and {\bf q} have axis $S$ and $R$ ($\hat{\sigma}_x$ and $\hat{\sigma}_z$), $S$ and $T$ ($\hat{\sigma}_x$ and $\hat{\sigma}_y$), and $T$ and $R$ ($\hat{\sigma}_y$ and $\hat{\sigma}_z$) as quantization axis, respectively. Therefore, to implement each step of DQW in kagome lattice should compose of evolutions along different axis in particular order depending on the initial position of the particle. For example, if the initial position is {\bf p} (as marked in Fig.\,\ref{fig:4}(b)), each step of DQW can be realized by $\hat{W}^{kag}(\theta) = \hat{W}^{kag}_{\sigma_x}(\theta)\hat{W}^{kag}_{\sigma_z}(\theta)\hat{W}^{kag}_{\sigma_y}(\theta)$.

\section{Discussion} 
\label{conc}
Motivated by the recent advancements in the theoretical  analysis and the experimental implementation of the DQW, we have presented the scheme for implementing DQW on different 1D, 2D, and 3D lattice structures using different Pauli basis states as the position translation states for each dimension. Our scheme using Pauli basis which was briefly presented for evolution on square lattice\,\cite{CBS10, CB13} has now been extended in this report for evolution on cubic, triangular, and kagome lattice. During the shift from one pair of the Pauli basis state as the position translation state to the other for the evolution on 2D and 3D lattice, an effect of inherent quantum coin operation of $\theta =\pi/4$ is seen. In various physical setting, it is quite common to use Pauli basis states as the position translation states to describe the dynamics, for example, in quantum optics and optical lattice \cite{DDL03}. This makes our scheme an experimentally realizable one in higher dimensions without the use of an extra resource to implement external quantum coin operation. However, an external coin operation can be used as an additional resource to tailor the dynamics. From this we can conclude that it can be straight away extended to other Bravais lattice and to higher dimensions by simply permuting the three Pauli basis states as position translation states for each evolution axis with the minimum resources. 
\par
The dynamics in many quantum systems are governed by the Hamiltonians and the Hamiltonian for the DQW was also recently obtained in the basis of Fourier modes and used to explore topological phases\,\cite{KRB10}.  In our approach, without changing to the Fourier modes, starting from the unitary operators used for the evolution we obtained three different forms of the Hamiltonian for walk on 1D  discrete-position space using different Pauli basis states as the position translation states and this was extended to different 2D and 3D lattices. In the unit position incremental setting on a line, square lattice and cubic lattice the differential operator form of the Hamiltonian was shown to be identical to the two-component Dirac Hamiltonian. On triangular lattice we showed that and Hamiltonian contains a second order deferential term deviating from the Dirac-like structure. Though the similarity of DQW with the Dirac equation was established earlier\,\cite{BB94, DM96, Str07, CBS10, GDB12, DDL03} using different approaches, the connection was at the continuum limit and restricted to 1D  for a very small or vanishing value of the quantum coin operation parameter $\theta$.  Very recently, when this manuscript was already in a complete form, a new result establishing the connection between the Dirac equation and quantum walk in higher dimension was reported\,\cite{ANF13}. Using different combination of techniques, the two-component Dirac equation is discretized to arrive at the two-state DQW in  1D and 2D position space. To establish the connection for the 3D space, they changed to four-component Dirac equation.  In this report our approach is different, we start from the two-state DQW on 1D, 2D, and 3D lattice and arrive at the two-component Dirac-like Hamiltonian for all value of $\theta$ (massless and massive case) as long as the evolution axis in the lattice are orthogonal to each other (line, square and cubic). However, both these works compliment each other by starting from the different initial from and establishing the connections in 1D and 2D position space. 
\par
This scheme for walk on different lattices and description of dynamics in Hamiltonian form helps to further explore topological phase, establish connection between physical process in nature which are generally not 1D and does not involve larger internal (more than two) dimension of the particle. Most importantly, the structural similarity DQW with the two-component Dirac Hamiltonian for particle in 1D,  2D and 3D space can lead to more intriguing studies on the relativistic effects in quantum walks and the role DQW can play to model and simulate the relativistic effects in the laboratory settings. The Dynamics in many of the quantum condensed matter systems are governed by the usual tight-binding Hamiltonian that describes the movement or the particle to a neighbouring site or by the Dirac-like Hamiltonian (for example, graphene).  This clearly establishes that DQW can be used to model the dynamics in a wide spectrum of physics systems governed by Dirac-like Hamiltonian to the relativistic dynamics.


\appendix
\section{The Hamiltonian and its differential form} 
\label{app1}
To obtain the effective Hamiltonian form, we will take the logarithm of  $\hat{W}_{\sigma_z}(\theta) = e^{-i \hat{H}_{\sigma_z}(\theta)}$,   
 \bea
\label{Ham}
-i\hat{H}_{\sigma_z}(\theta) &=& \ln \begin{bmatrix}
   \mbox{~~}\cos(\theta) e^{-i\hat{P}_z}     &     &    \sin(\theta)e^{-i\hat{P}_z}
  \\ - \sin(\theta)e^{+i\hat{P}_z}  & &  \cos(\theta)e^{+i\hat{P}_z} 
\end{bmatrix}  
=  \ln \,(\hat{A} )  = \hat{V} \,\ln(\,\hat{\lambda}_{\sigma_z}\,)\, \hat{V}^{-1}.
\eea
Here 
$\hat{\lambda}_{\sigma_z} =\begin{bmatrix}
\hat{\lambda}^-_z & & 
0 \\
0 & & \hat{\lambda}^+_z \end{bmatrix}$
with 
$\lambda^{\mp}_z = \cos(\theta) \cos(\hat{P}_z) \mp \sqrt {\cos^2(\theta)\cos^2(\hat{P}_z) - 1}
~= ~ \cos(\hat{\omega}_z) \mp i\sin(\hat{\omega}_z) ~=~ e^{\mp i\hat{\omega}_z}$.
The matrix 
\bea
 \hat{V} =\frac{1}{\sin(\theta) e^{i \hat{P}_z}} \begin{bmatrix}
 \cos(\theta)e^{i \hat{P}_z} - e^{-i\hat{\omega}} & &  \cos(\theta)e^{i \hat{P}_z} - e^{i\hat{\omega}} \\
\sin(\theta) e^{i \hat{P}_z}  & & \sin(\theta) e^{i \hat{P}_z} \end{bmatrix}~~
~;~
\hat{V}^{-1} =  \frac{1}{2i \sin(\hat{\omega}_z)}  \begin{bmatrix}  
~\sin(\theta) e^{i\hat{P}_z}  & & 
e^{i\hat{\omega}} - \cos(\theta) e^{i\hat{P}_z} \\
-\sin(\theta)e^{+i\hat{P}_z} & &
- e^{-i\hat{\omega}} + \cos(\theta) e^{i\hat{P}_z}\end{bmatrix} ~~
\eea
are composed of eigenvectors of $\hat{A}$ and its inverse, respectively. By substituting these elements into Eq.\,(\ref{Ham}) and simplifying we obtain,
\bea
\label{sub1}
\hat{H}_{\sigma_{z}}(\theta)= \frac{\hat{\omega}}{\sin(\hat{\omega})} 
\begin{bmatrix}
\cos(\theta) \sin(\hat{P}_z)  & 
i\sin(\theta)e^{-i\hat{P}_z} \\
-i \sin(\theta)e^{i\hat{P}_z} 
&
-\cos(\theta) \sin(\hat{P}_z) 
\end{bmatrix} 
= \frac{\hat{\omega}}{\sin(\hat{\omega})} 
\begin{bmatrix}
\cos(\theta) \sin(\hat{P}_z)  & 
-i\sin(\theta)e^{-i\hat{P}_z} \\
-i \sin(\theta)e^{i\hat{P}_z} 
&
\cos(\theta) \sin(\hat{P}_z) 
\end{bmatrix} \cdot \sigma_z.
\eea
The differential form of the preceding expression in the unit spatial incremental setting can be obtained by analyzing the effect of the operators $e^{\pm i\hat{P}_z}$ and $\sin({\hat{P}_z})$ in the matrix on the state $\psi^{\downarrow (\uparrow)} (z, t) = \psi (z, t)$,
\bea
\sin({\hat{P}_z})  \psi (z, t) = i\Bigg ( \frac{ e^{-i\hat{P}_z} - e^{i\hat{P}_z} }{2} \Bigg ) \psi (z, t) = \frac{i}{2} \Big ( \psi (z-1, t) - \psi(z+1, t) \Big ) = -\frac{i}{2} \Big ( \psi (z+1, t) - \psi(z-1, t) \Big ).
\eea
Writing the central difference form in the preceding expression as differential operator we obtain,
\bea
\label{sub2}
\sin({\hat{P}_z}) \psi(z, t) &\approx& -i \frac{\partial}{\partial z} \psi (z, t)  \implies \sin({\hat{P}_z}) \approx -i \frac{\partial}{\partial z}   \equiv \hat{P}_z.
\eea
Similarly,
\bea 
\label{sub3}
e^{\pm i\hat{P}_z}\psi (z, t) = \psi (z\pm 1, t)  = \psi (z \pm 1, t) - \psi (z, t)  + \psi (z, t)  \approx \Big(\pm \frac{\partial}{\partial z} +1 \Big )\psi (z, t)  \implies e^{ \pm i\hat{P}_z}  \approx \Big(\pm \frac{\partial}{\partial z} +1 \Big ).
\eea
Ignoring the higher order terms in the expansion of $\sin(\hat{\omega})$  we get, $\frac{\hat{\omega}}{\sin(\hat{\omega})} \approx 1$. Therefore, substituting Eq.\,(\ref{sub2}) and \,(\ref{sub3}) into Eq.\,(\ref{sub1})  we obtain the differential form of the Hamiltonian operator which is identical to the two-component Dirac Hamiltonian.
\par
For evolution with basis states of $\hat{\sigma}_x$ and $\hat{\sigma}_y$ Pauli  operators as position translation states, the same procedure can be used to obtain Eqs.\,(\ref{hamil2b}) and \,(\ref{hamil2c}) by replacing $\hat{P}_z$ by $\hat{P}_x$ and $\hat{P}_y$, respectively.  Here using $j \in \{x, y \}$ we arrive at the differential form of the operators which upon substitution we obtain an Hamiltonian given by  Eqs.\,(\ref{hamil2ba}) and \,(\ref{hamil2ca}) which are structurally similar to the two-component Dirac Hamiltonian. One of the useful relation to arrive at the differential operator form is, 
\bea
\Big (\cos(\theta)\sin(\hat{P}) \mp i \sin(\theta)\cos(\hat{P}) \Big ) \psi (j, t) &=& \frac{i}{2} \Bigg[ \cos(\theta) \Big [
\psi (j-1, t) - \psi (j+1, t) \Big ] \mp \sin(\theta)\Big[ \psi (j-1, t) + \psi (j+1, t) \Big ] \Bigg ] \nonumber \\
&=& \frac{i}{2}\Big[ \big ( \cos(\theta)  \mp \sin(\theta) \big)  \psi (j-1, t) - \big ( \cos(\theta)  \pm \sin(\theta)\big )  \psi (j+1, t)\Big ], 
\eea
adding and subtracting the RHS by $\frac{i}{2} \big( \cos(\theta)  \mp \sin(\theta) \big ) \psi(j, t) + \frac{i}{2} \big ( \cos(\theta)  \pm \sin(\theta) \big ) \psi(j, t)$ we get a difference form which can be approximated to the differential operator such that,
\bea
\cos(\theta)\sin(\hat{P}) \mp i \sin(\theta)\cos(\hat{P}) 
 &\approx& \frac{i}{2} \Big [ \big ( \cos(\theta) \mp \sin(\theta)\big ) \Big [ - \frac{\partial}{\partial j} +1 \Big ]  -\big ( \cos(\theta) \pm \sin(\theta) \big ) \Big [  \frac{\partial}{\partial j} +1 \Big ] \Big ], \nonumber \\
 &=& i \Big [ - \cos(\theta) \frac{\partial}{\partial j}  \mp \sin(\theta)  \Big ].
 \eea
\section{Differential form of the Hamiltonian for evolution on triangular lattice} 
\label{app2}
For triangular lattice $\hat{P}_z$ in Eq.\,(\ref{hamil2a}) is replaced by $2\hat{P}_{x}$,  $\hat{P}_x$ in Eq.\,(\ref{hamil2b}) is replaced by $\hat{P}_{x} + \hat{P}_{y}$ and 
$\hat{P}_y$ in Eq.\,(\ref{hamil2c}), is replaced by  $\hat{P}_{x} - \hat{P}_{y}$.  Below we obtain the differential form of the operators in finite position incremental setting, 
\begin{align}
\sin(2{\hat{P}_x})  \psi (x,y, t) &= \frac{i}{2} \Big [ \psi(x-2, y, t)  -\psi(x+2, y, t)  \Big ]   \approx - 2 i  \frac{\partial }{\partial x} \psi(x, y, t), \nonumber \\
e^{\pm 2i \hat{P}_x}\psi (x,y, t) &\approx \Big ( \pm \frac{\partial}{\partial x} +1 \Big ) \psi (x+1, y, t) \approx \Big ( \frac{\partial^2}{\partial x^2}  \pm 2  \frac{\partial}{\partial x} +1 \Big ) \psi (x, y, t), \nonumber \\
\implies  \sin(2{\hat{P}_x}) &\approx - 2 i  \frac{\partial }{\partial x}    ~~~;  ~~~ e^{\pm 2i \hat{P}_x} \approx \Big ( \frac{\partial^2}{\partial x^2}  \pm 2  \frac{\partial}{\partial x} +1 \Big ).
\end{align}
Considering $\frac{\hat{\omega}}{\sin(\hat{\omega})} \approx 1$ and substituting the differential form of the operators we obtain, 
\bea
\label{hamiltriaa}
\hat{H}_{\sigma_{z}}^{tri}(\theta)=  
-i \Bigg ( \begin{bmatrix}
 \cos(\theta)  & 
\sin(\theta)   \\
 \sin(\theta)   
&
- \cos(\theta) 
\end{bmatrix}   \cdot 2 \frac{\partial }{\partial x}
+ \begin{bmatrix}
~~0  & &
-\sin(\theta)    \\
\sin(\theta)    
& &
0 
\end{bmatrix} \cdot  \frac{\partial^2}{\partial x^2}  \Bigg )
+  \begin{bmatrix}
~~0  &  &
i   \\
-i 
& &
0
\end{bmatrix}  \sin(\theta).
\eea
Using the relations, 
\begin{align}
\sin(\hat{P}_x + \hat{P}_y)    &\approx -i \Big( \frac{\partial}{\partial x}  + \frac{\partial}{\partial y} \Big )~~~~ \mbox{and}  \nonumber \\
\cos(\theta)\sin(\hat{P}_x + \hat{P}_y) \mp i \sin(\theta)\cos(\hat{P}_x + \hat{P}_y)  &\approx -i\cos(\theta) \Big( \frac{\partial}{\partial x}  + \frac{\partial}{\partial y} \Big ) \mp i\sin(\theta) \Big ( \frac{\partial^2}{\partial x \partial y} + 1 \Big ),
\end{align}
\bea
\label{hamiltribb}
\hat{H}_{\sigma_{x}}^{tri}(\theta)= -i \Bigg (
\begin{bmatrix}
 \sin(\theta)    \, & 
~~ \cos(\theta) \, \\
 \cos(\theta)  \,
&
- \sin(\theta) \end{bmatrix} \cdot \frac{\partial}{\partial x} +
\begin{bmatrix}
 \sin(\theta)    \, & 
~~ \cos(\theta) \, \\
 \cos(\theta)  \,
&
- \sin(\theta) \end{bmatrix} \cdot \frac{\partial}{\partial y} +
\begin{bmatrix}
0 \, & 
 -i\sin(\theta)    \, \\
i\sin(\theta) \,
&
0
\end{bmatrix} \cdot \frac{\partial^2}{\partial x \partial y} \Bigg )
+  \begin{bmatrix}
0    &
-i  \\
i 
 &
~~~0
\end{bmatrix}  \sin(\theta).~~
\eea
Using the relations, 
\begin{align}
\sin( \hat{P}_x - \hat{P}_y)   &\approx -i \Big( \frac{\partial}{\partial x}  - \frac{\partial}{\partial y} \Big ) ~~~~\mbox{and}  \nonumber \\  
\cos(\theta)\sin(\hat{P}_x - \hat{P}_y) \mp i \sin(\theta)\cos(\hat{P}_x - \hat{P}_y) &\approx
-i\cos(\theta) \Big ( \frac{\partial}{\partial x}  - \frac{\partial}{\partial y}  \Big )
 \pm i\sin(\theta) \Big ( \frac{\partial^2}{\partial x \partial y} - 1 \Big ),
\end{align} 
\bea
\label{hamil2ctri}
\hat{H}_{\sigma_{y}}^{tri}(\theta) =   -i \Bigg (
\begin{bmatrix}
~~\sin(\theta)    &
~i\cos(\theta)   \\
-i\cos(\theta)  &
- \sin(\theta) 
\end{bmatrix}  \cdot \frac{\partial}{\partial x}+ 
\begin{bmatrix}
- \sin(\theta)  &
-i\cos(\theta)   \\
~i \cos(\theta)   &
~~~ \sin(\theta) 
\end{bmatrix} \cdot \frac{\partial}{\partial y}+ 
\begin{bmatrix}
0    &
i \sin(\theta)  \\
- i \sin(\theta)  &
0  
\end{bmatrix} \cdot \frac{\partial^2}{\partial x \partial y} \Bigg ) + 
 \begin{bmatrix}
0    &
-1   \\
1 
 &
~~~0
\end{bmatrix}  \sin(\theta).~~
\eea


\vskip 0.3in

{\bf Author Contributions:} \\
 CMC designed the scheme, derived the expressions, checked the numerics, prepared the figures and wrote the manuscript.

\vskip 0.3in

{\bf Competing financial interests :}\\
The author declare no competing financial interests.

\end{document}